\documentclass[10pt,preprint]{aastex}

\singlespace

\newcommand\hb{H${\beta}$~}
\newcommand\ha{H${\alpha}$~}
\newcommand\nii{[N{\sc ii}]~}
\newcommand\civ{[C{\sc iv}]~}
\newcommand\sii{[S{\sc ii}]~}
\newcommand\siii{[S{\sc iii}]~}
\newcommand\oii{[O{\sc ii}]~}
\newcommand\oi{[O{\sc i}]~}

\newcommand\oiii{[O{\sc iii}]~}

\newcommand\sm{M$_{\odot}$}
\slugcomment{}
\shortauthors{Stanghellini et al.}
\shorttitle{SMC Planetary Nebula morphology and evolution} 

\begin{document}

\title{{\it Space Telescope Imaging Spectrograph} slitless observations of Small Magellanic
Cloud Planetary Nebulae: a study on morphology, 
emission line intensity, and evolution.
\footnote{Based on observations made with the NASA/ESA Hubble Space Telescope,
obtained at the Space Telescope Science Institute, which is operated by the 
Association of Universities for Research in Astronomy, Inc., under NASA 
contract NAS 5--26555}}

\author{Letizia Stanghellini\altaffilmark{2}}
\affil{Space Telescope Science Institute, 3700 San Martin Drive,
Baltimore, Maryland 21218, USA; lstanghe@stsci.edu}

\author{Richard A. Shaw}
\affil{National Optical Astronomy Observatory, 950 N. Cherry Av.,
Tucson, AZ  85719, USA; shaw@noao.edu}

\author{Bruce Balick}
\affil{Department of Astronomy, University of Washington, Seattle,
Washington 98195; balick@astro.washington.edu}

\author{Max Mutchler}
\affil{Space Telescope Science Institute; mutchler@stsci.edu}

\author{J. Chris Blades}
\affil{Space Telescope Science Institute; blades@stsci.edu}

\and

\author{Eva Villaver}
\affil{Space Telescope Science Institute; villaver@stsci.edu}

\altaffiltext{2}{Affiliated with the ESA Space Telescope Division; 
on leave from INAF- Osservatorio Astronomico di Bologna}

\begin{abstract}
A sample of 27 Planetary Nebulae (PNs) in the Small Magellanic Clouds (SMC) have been observed
with the Hubble Space Telescope Imaging Spectrograph ({\it HST}/STIS) to determine their morphology,
size, and the spatial variation of the ratios of bright emission lines. 
The
morphologies of SMC PNs are similar to those of LMC and Galactic PNs.
However, only a third of the resolved SMC PNs are asymmetric,
compared to half in the LMC. The low metallicity environment of the 
SMC seems to discourage the onset of bipolarity in PNs.
We measured the line intensity, average surface brightness (SB), and photometric radius
of each nebula in \ha, \hb, \oiii $\lambda$4959 and 5007, \nii $\lambda$6548 and 6584, 
\sii $\lambda$6716 and 5731, He I $\lambda$6678, and \oi $\lambda$6300 and 6363.
We show that the surface brightness to radius relationship is
the same as in LMC PNs, indicating its possible use as a distance scale  
indicator for Galactic PNs. 
We determine the electron densities and the ionized masses of the nebulae where
the \sii lines were measured accurately, and we find that the SMC PNs are denser than the
LMC PNs by a factor of 1.5. 
The average ionized mass of the SMC PNs is 0.3 \sm.
We also found that the median \oiii/\hb 
intensity ratio in the SMC is about half than the corresponding LMC median.
We use Cloudy to model the dependence of the \oiii/\hb ratio 
on the oxygen abundance. Our models encompass very well the 
average observed physical quantities. We suggest that 
the SMC PNs are principally cooled by the carbon lines, making it hard to
study their excitation based on the optical lines at our disposal. 
 
\end{abstract}


\vfill
\eject

\section{Introduction}

To understand the final phases of stellar evolution of intermediate mass stars,
one has to study the interplay of the stellar and nebular components. During 
its lifetime, and especially toward the end of its life, a star of mass
between 1 and 8 \sm~ experiences extreme mass loss, and nuclear burning episodes
followed by the occurrence of the dredge-ups that change the chemical mix of its envelope.
A planetary nebula is formed by the ejected envelope of its central star
at the tip of the AGB, and it can be shaped
in part by a fast wind from the remnant, during its subsequent evolution.

The process of PN formation and evolution has been studied
both theoretically and observationally for decades, yet many fundamental questions
remain only partially solved. For example, what is the origin of the different 
morphologies, and what phenomenon, or series of phenomena, decide the final
PN morphology? Does progenitor mass alone dictate 
the dynamic evolution rate and expansion speed? What is the connection
between binary star evolution and the physical properties of PNs (chemistry, morphology,
etc.)? What is the role of the interstellar medium (ISM) on PN structure? How much is the 
nebular gas mix contaminated by the interaction with the
interstellar medium? To what extent do PNs enrich C, N, and O in the ISM? 

A number of studies based on the comparison of PN and stellar properties with 
models have shed light on some of these and other questions. 
Even so,
observations of planetary nebulae and their central stars are notoriously difficult to 
compare with adequate theoretical models for a variety of reasons. First, the
stellar and nebular components are often observed and analysed 
separately, yet they are of common origin; the nebula
carries, in its chemical composition, the nuclear burning and dredge up history of
the stellar progenitor, and its morphology carries the signature of the
wind history of its central star. Second, the inferred
evolutionary times of the stellar and nebular component 
within the post-AGB phase are unsynchronized. Although theoretical models have been 
developed to explain the nuclear evolution of the AGB and post-AGB phases, 
mass loss rates and geometries of the envelope ejection have 
not yet been 
successfully modeled from the
stellar viewpoint. Dynamic theoretical models have explained the observed nebular
morphology only if ad hoc assumptions are made about 
the evolution of the central star. Photoionization models 
are able to predict the strength of optical and UV emission lines that 
appear during the evolution of PNs, 
but once again without yielding much insight into the numerous details of morphology,
the evolution of the progenitor star. We believe that only with large samples of well defined,
high quality data sets that allow the 
identification and significance of the crucial nebular and stellar physical parameters can one make
steps forward in this field of study. Large amounts of data of high quality
will be able to constrain the stellar and nebular models, so that 
open questions on PN formation and evolution may be addressed.

In the quest for acquiring the best observational sets for model testing, 
PNs in the Magellanic Clouds (SMC, LMC) deserve special attention. LMC and SMC PNS
have known distances 
and low field reddening, in contrast with Galactic PNs. 
Our aim in the last few years has been that of acquiring and analyzing 
a large sample of LMC and SMC PNs spectra and images, that, together with the 
data existing in the literature, permits the construction of an ideal data set  
for model testing, and for comparison with disk and halo Galactic PNs.

The importance of observing and classifying SMC PNs morphology, 
and discerning the evolutionary
stage of their central stars, is multifold. First, the ambient metallicity
of the SMC is closer to that of a primordial galaxy than the LMC or the Galaxy
are, thus making the SMC an unique astrophysical laboratory to study
metal depleted yet resolved stellar populations. 
Second, by observing a sizable sample of SMC PNs with {\it HST} (i.e., resolving
their morphology and size) we will be able to study the dependence between morphology
and abundances in a low metallicity environment. 

The Space Telescope
Imaging Spectrograph (STIS) has proven to be a very efficient way to 
acquire broad band and monochromatic images of PNs smaller than about 2 \arcsec. 
In two previous papers of this series
\citep{sha01,sta02} we have presented the results from the STIS slitless
spectroscopy of the first set of 29 LMC PNs.
In this paper, we present the results relative to the observations of 27 SMC PNs 
({\it HST} 
program No. 8663). In $\S$2 we discuss the observation strategy and the data analysis of this
particular observing set; we include there the rationale for target selection, the data
calibration and slitless spectral extraction, the spectral analysis, the 
photometric radii determinations, and the morphological classification.
In $\S$3 we present the scientific results derived from this set of images and spectra,
including the discussion
of the line intensity, the surface brightness in the different lines, the excitation of the
nebulae, and their optical extinction, all discussed across
morphological classes. Section 4 discusses the results so far and the future
endeavors
in this study. The analysis of the central stars of SMC PNs from the images and spectra 
of the data set of Program 8663 will be discussed in a separate paper.

\section{Observations, data calibration, and analysis}

\subsection{Target selection, and misclassifications}

Our list of 55 targets was compiled with the intent to obtain data of
all known SMC PNs, with accurate coordinates. Most targets came
from the \ha~ survey catalog by \citet{ma93}. 
The majority of these PNs
are bright enough to be observed in snapshot mode (less than 1 {\it HST}
orbit). We included in the
target list three objects that have been previously observed with the 
Wide Field and Planetary Camera 1, and with the Faint Object Camera. 
This was done to check the consistency of the morphological classification
obtained with archived instruments and the STIS.

Our {\it HST} observations were
performed in snapshot mode. The targets were selected from the original list 
by the scheduling specialists at STScI, accordingly to the available
observing slots. Typically, shorter exposure (i.e., brighter targets) are
favored.
The program was completed at the 53 $\%$ level, which is normal for a
successful snapshot proposal. Among the 29 targets observed (Table 1), 
27 are the SMC PNs analyzed in this paper. Two (MG~2 and Ma~1796)
are misclassified H II regions, whose images and analysis 
will be published in another paper.
As it turned out, faint targets are slightly underrepresented in our sample.

\subsection{Observing technique}

The observations were acquired with STIS. 
All observations were made with the CCD detector, in direct imaging
(50CCD)
and slitless mode. The spatial scale of the CCD is 0.051 arcsec pixel$^{-1}$,
corresponding to 0.0144 pc pixel$^{-1}$
at the distance of the SMC. This 
allows a very good spatial resolution to study PN morphology 
and size, comparable to what used in ground-based imaging observations 
of Galactic PNs.

Each imaging observation was split in two, to
allow easy cosmic rays removal. The slitless spectra were acquired
with the G430M and G750M gratings for most nebulae. 
Observations with the G430M grating cover the range 4818 \AA~ to
5104 \AA~ at 0.28 \AA~ pixel$^{-1}$, and those with the G750M grating cover 
the range 6295 \AA~ to 6867 \AA~ at 0.56 \AA~ pixel$^{-1}$.
The exposures where planned to have a good
signal-to-noise ratio in the \oiii $\lambda$5007 and \ha lines. Several
additional lines have been detected in many PNs, including \hb and
\oiii $\lambda$4959 using G430M, and \oi $\lambda\lambda$6300, 6363, \siii $\lambda
6312$, \nii $\lambda\lambda$6548, 6584, He I $\lambda$6678, and \sii 
$\lambda\lambda$6716, 6731 using G750M. In some cases the G430M 
exposures were skipped, limited by the {\it HST} snapshot duration.

The observing log is reported in Table 1, where we list the targets, the
observing date, the
data set name, the spectral element used in the observations, and the 
exposure time and number of exposures obtained. Aliases are identified in the 
table notes.
The STIS data were calibrated using the standard
pipeline system, as in the LMC data \citep{sha01}.

Figures 1 through 6 show the observed SMC PNs in the three
observing modes: left panels show the broad band CCD images; central panels
show the \ha and \nii images; right panels show, when available,
the \oiii 5007 \AA~ images. 

\subsection{One dimensional spectral extraction, and line intensities}

Spectral analysis of the SMC PNs have been performed similarly to
that of LMC PNs \citep{sta02}. For most PNs observed, the combination of
dispersion and spatial scale allows a clear separation
of the monochromatic images for all emission lines. 
Exceptions are J~27, where broad and monochromatic images are at the limit of
detectability, and MA~1682, where the \nii and \ha images may have partial
overlap. No images are severely overlapped, thus the one dimensional spectral
extraction was adequate for nebular line flux and ratio analysis. We extract the
one-dimensional spectra and applied a photometric calibration 
using the standard STIS calibration pipeline module {\bf x1d} 
\citep{McGrath_etal99}.
We used extraction boxes for the nebulae large enough to encompass all the
nebular features, but snug enough as to exclude most of the sky background
from the extraction. Sky background regions were selected for each object to avoid
stray stellar photons from field stars. The background was then averaged and
subtracted. 

We measured emission line intensities with IRAF\footnote{IRAF is distributed by
the National Optical Astronomical Observatory, which is operated by the 
Association of Universities for research in Astronomy, Inc., under cooperative
agreement with the National Science Foundation.} {\bf splot} task, 
fitting gaussians to individual lines, while estimating the continuum level. 
In the cases in which the emission lines were notably non gaussian, we 
estimated the line flux as measured from the area above the continuum.

In Table 2 
we report the measured line intensities for the SMC PNs in 
our sample. Column (1) gives the common names; column
(2) gives the logarithmic \hb intensities, not corrected for extinction, in 
erg cm$^{-2}$ s$^{-1}$; column (3) lists the logarithmic optical extinction
at \hb \citep{ost89}; columns (4) to (14) give the line intensities for each nebula,
relative to \hb=100, not corrected for extinction. Line identification are
given at the heads of the columns.

From the analysis of our spectral line measurements, and 
given the similarities of this data set to the ones analyzed in
\citet{sta02}, we confirm that the errors in the line intensities of
Table 2 ($\delta {\rm log} F$) are (in dex): $\delta {\rm log} F < .05$ if 
${\rm log} F > -12.25$;  $\delta {\rm log} F<.15$ if $-12.25 > {\rm log} F > -12.75$;
$\delta {\rm log} F < .2$ if $-12.75 > {\rm log} F >-13.5$; $\delta {\rm log} F< .25$
if $-13.5 > {\rm log F} > -14.5$ ; and $\delta {\rm log} F<.55$ if ${\rm log} F < -14.5$.

In Figure 7 we compare the measured line intensities 
with those available in the literature. We used the data from selected references
\citep{dop90,dop91a,dop91b,vdm}. 
Some authors give the intensity corrected for extinction, so before comparison to
the present results we
un--correct the intensity ratios by using the extinction constant given in the same reference. 
We have comparisons for several emission lines in MG~8, MG~13, SMP~1,
SMP~6, SMP~11, SMP~13, SMP~14, SMP~17, SMP~25, and SMP~26. A comparison of our fluxes to
with previous values (Fig. 7) 
shows a generally good agreement. The correlation coefficient between the two sets of 
fluxes is 0.994, with RMS scatter of 0.2 dex.
Uncertainties in the reference fluxes are typically quoted 20$\%$ to 50$\%$ for high and
low fluxes respectively, while our errors are generally smaller \citep{sta02}. The outlying point in
Figure 7 corresponds to the \nii $\lambda$ 6584 \AA~ flux measurement of target SMP 17. We believe that our 
measurement is correct, since the reddened
\nii $\lambda$ 6584 to $\lambda$ 6548 intensity ratio is 2.1, 
while the reference does not detect the \nii $\lambda$6548 \AA~ line.

\section{Results}

\subsection {Dimensions and morphology of SMC planetary nebulae} 

In Table 3, columns 2 and 3, we give 
the positions of the central stars of the nebulae from the continuum images, where
observed. Alternately, we give the geometric center positions of the 
PNs. Columns (4) and (5) of Table 3 give the photometric radii of the 
nebulae, measured as described in \citet{sta99}, and the nebular dimensions,
measured from the 10$\%$ brightness contour.

Figures 1 through 6 show the SMC PNs 
images, exhibiting the same range of morphological types to those of LMC
and Galactic PNs. 
The morphological classification used in this paper is the same as
that in the other papers of this series \citep{sha01,sta02}: 
we classify PNs as Round, Elliptical, Bipolar
Core, Bipolar, and Pointsymmetric. We distinguish between Round and
Elliptical if the axial ratio is larger than unity by at least 10 percent.
Bipolarity follows the classic definition \citep{sta93}, while bipolar core 
is defined in \citet{sta99}. Note that bipolar core Pns of this sample are
defined as such only if their BC is apparent above the 10 percent intensity
contour.
Morphology is based primarily on the \ha images, which differs only rarely
from that of the \oiii $\lambda$5007 images \citep{man96a}. 
Column (6) of Table 3 gives the morphological
classification. In a couple of cases, we use the terminology of Round and Elliptical PNs with
inner structures. These structures may
be bipolar cores, although we believe that we have the
necessary spatial resolution to make the distinction. Question marks indicate that the 
morphological determination is ambiguous or uncertain.
In the following paragraphs, we
describe in detail the morphological types of our sample
PNs. 

\begin{itemize}

\item{J~4 (E): the field is crowded, making the ID on the
clear image a little uncertain. The spectrum is
reasonably well exposed, and shows emission in the full set of
emission lines. The nebular morphology is elliptical. We detected extremely faint
emission in the \oiii frame up to 1\arcsec.42 from the center of the PN in the 
spatial direction.}

\item{J~18 (R): This PN is round in \ha and has an elliptical shape in \nii. It 
does not show a bipolar core. The \nii $\lambda$6584 emission is brighter than \ha.}

\item{J~23: This object is apparently spatially unresolved, and its identification 
on the clear image is difficult. 
The stellar spectrum shows strong and broad \ha, and a weaker $\lambda$6584 \nii line. 
We also detect He I $\lambda$6678. 
The \ha velocity width appears to
be about 200-300 km/s, which is broad for a PN. 
This object could not be
a common symbiotic star, it would have a giant companion with an unusually 
bright \ha emission line 
and a faint underlying continuum \citet{1989ApJ...339..844B}.}

\item{J~27 (B?): This object is roughly box-shaped, and may be the remnant of a bipolar. 
The
nebula is barely detected in the broad-band exposure, at roughly 20 counts/pixel
above the background. The spectrum shows faint \nii emission, with some hint of
\ha emission. No central star is detected.}
    
\item{MA~1682 (B): This is an extreme bipolar PN, with a prominent central torus seen
edge on. The \ha and [N
II] lines are very strong in the spectrum, and \sii is barely detected; no other lines
are detected. We don't have a blue spectrum. The central star (CS)
is easily detected in the continuum image.}
    
\item{MA~1762 (E(bc)): This is an elliptical PN, with a bi-nebulous inner core. We detected 
\ha and marginally \oi $\lambda$6300. The CS is easily detected in the
continuum image, and in the spectrum.} 
    
\item{MG~8 (E): This elliptical PN has a very distinct ring structure, with a hint of ansae 
(or arms) that extend
~0.1 \arcsec~along the major axis from the 10$\%$ contour. The arm-like structure leads
to a pointsymmetric classification, but to be conservative we classify this elliptical,
possibly bipolar core.
\ha is very strong, and
\nii is apparent but weaker, \sii is very weak, and we could also detect
a trace of the \oi $\lambda$6300 emission. 
\hb is well exposed, and \oiii is only a bit stronger. The
morphology in \oiii differs in detail from the \ha and other line
morphology, in that the emission comes primarily from
a ring that is inside the emission from most other lines.
The CS is exceptionally bright in the continuum image; could it be a 
binary companion to the true CS?}
    
\item{MG~13 (E): This elliptical PN has a very knotty ring structure, and is
classified as E(s). \ha is very strong, but neither \nii nor any
other line is detected in the G750M spectrum. The CS is easily detected in the
continuum image.}
       
\item{SMP~1: Marginally resolved round or elliptical PN with no detected image
of the central star. 
Most emission lines are present,
and there may be a trace of He I at 4927 \AA~ as well. The \sii lines are marginally
detected.}
    
\item{SMP~6 (E): Elliptical PN, the \nii morphology is slightly different from that
of the other lines, in that it shows a larger elongation. }  

\item{SMP~8 (R): This Round PN has very strong \ha and \oiii emission,
with \nii and He I at the ~1$\%$ level, and \sii weaker still. \oi is
marginally detected at 6300 \AA. The CS is detected in the continuum image, 
but there is a significant contribution from the nebular emission.}

\item {SMP~9 or J~3 (R): This round PN has a bright hemisphere of emission to the NE, 
making the whole somewhat asymmetric in appearance. We are not sure that the inner
structure clearly discloses a bipolar core, but there is a lot of structure.
\ha and \oiii are very strong, but \nii is
fairly weak. The asymmetry is pronounced in the \nii lines.
\sii and He I,
are also present but very weak, and \oi $\lambda$6300 may be marginally detected. The CS
has been detected in the continuum image.}

\item{SMP~11 or J~8 (B): This complex bipolar PN has a distinct dark band that divides the 
bright core into two,
unequal lobes. This looks very much like a case of a strong bipolar ring, viewed some
distance from the plane. A faint bipolar structure is even evident up to 2 \arcsec~
from the center (particularly to the south), but the ribbed, fan-like structure is
projected off-center on the sky, and \nii $\lambda$6548 is blended with the \ha
gaussian wing. \ha and \oiii are very strong, but \nii $\lambda$6584 is
fairly weak; these lines all show the same morphological structures. \sii and He I,
are also present but very weak, and \oi $\lambda$6300 is also detected. The CS
is detected in the continuum image, and possibly also in the G750M spectrum.}

\item{SMP~12 (E): Elliptical PN with detected central star. 
The central emission shows some structure (Es).
In the \oiii image there is a hint of extended lobe to the east.  }  

\item{SMP~13 or J~11 (R): the appearance of this PN is nearly stellar 
in the continuum,
but appears very mildly elliptical in the ionization lines. We classify it as round,
since the ellipticity is below the 10$\%$ level. Faint emission in the 
\oiii image is detected up to 1.42\arcsec~from the geometrical center. \ha and \oiii
are very strong, \nii, He I, and \sii are very weak but detected. Also weak are
\oi $\lambda\lambda$6300, 6363, and [S III] $\lambda$6312. The CS may be detected in the continuum image.}

\item{SMP~14 (R): This round PN has a distinct structure in the
inner core and faint ansae that extend
~0.3\arcsec~ from the 10$\%$ contour. \ha is very strong, and \nii is detected in the
spectrum, but is very weak. \sii and He I, are also present but weak. \oiii is quite
strong. The CS is easily detected in the continuum image.} 
    
\item{SMP~17 (E): This is an elliptical PN, with a very faint outer halo that 
is detected in the continuum
image, in \ha, and in \oiii at the 0.4$\%$ contour level, out to about 1.3\arcsec. 
the inner portion is
somewhat elliptical in all the observed lines except \nii, where it is ring-like.
the \oiii emission shows a marginal bipolar core.
\ha is very strong, and \nii is present but very weak. The CS is detected in the
continuum image, but there is a significant contribution from the nebular emission.}
    
\item{SMP~18 or J~19: This small PN with bright \ha and \oiii emission. The \nii
extension is smaller than in the other lines. It is barely resolved when compared to
the broad band images of the stars.}

\item{SMP~19 or J~20 (R): morphologically similar to J~3, the inner parts of this 
PN are slightly distorted in the \oiii lines and \ha. We classify it as R with structures. There
is outer emission of 1\arcsec~ radius.}

\item{SMP~20: This PN is bright but unresolved. \ha and \oiii are very strong, with \nii at
the 1$\%$ level, and He I, \sii, and \oi weaker still. The CS is detected in the
continuum image, but there is a significant contribution from the nebular emission.
The stellar spectrum shows a feature at about 6577 \AA, which is also seen in a few other targets in our sample.}
    
\item{SMP~22 (B?): This PN has a very interesting box/ring shape, with a possible emission
``arm'' toward the east, which is very evident in the \nii
images (could it be part of the ISM?). It is classified as bipolar. There are at least three 
high emission features within the ring. The line profiles of  all lines but \ha and
\hb are split, showing the ring morphology. }

\item{SMP~23 or J~26 (E(bc)): Elliptical PN with a bipolar core.}

\item{SMP~24 (E): Elliptical PN in the \ha and \oiii emission lines, it shows some
small scale structures in the light of \nii. This nebula may have a faint halo.}  

\item{SMP~25 (E): Elliptical Planetary nebula with likely broad stellar \ha emission line.}  

\item{SMP~26 (P): This PN could be pointsymmetric,
with two arms that extend from the NE and
SW of the center. It is possible that this PN is elliptical with low
ionization ansae. \ha is fairly strong, but \nii is nearly as bright, and shows
the arms much more clearly. The line profiles in the 1D spectra are split
in all lines except \ha. No CS is detected in the continuum image. }
    
\item{SMP~27 (R): This is a (very low ellipticity) elliptical
PN. There is a hint of structure in the \nii images only. \ha and \oiii
are very strong, with \nii at the 1$\%$ level, and He I, \sii, and \oi weaker still.
The CS is detected in the continuum image, but there is a significant contribution
from the nebular emission. }
    
\item{SP~34 (R): This is a round PN, with a faint outer halo that is detected in the 
continuum image and in \ha at the 5 $\%$ contour level. 
The inner portion shows significant
asymmetry in all the observed lines, in that the western edge is much brighter.
\ha is fairly strong, and \nii is present but much weaker. No CS is detected in
the continuum image. }
\end{itemize}

\subsection{Targets in common with prior {\it HST} programs}

We included in our STIS target list three objects that were previously observed with
the Planetary Camera 1 and the Faint Object Camera.
The rationale to include a few repeat targets from the sample of
\citet{sta99} was to check the morphological types and dimensions with both
methods and to asses the reliability of the pre-COSTAR archival data. 
All three were observed before the first {\it HST} servicing mission, thus
the images suffer for the uncorrected spherical aberration of the telescope mirror.
In an {\it HST} archival study \citep{sta99} it was found that morphology was 
well determined with the archived instruments, while dimensions of the nebulae with the 
photometric method were not always reliable because of the 
possible presence of fainter extended halos around the images.

Since program No. 8366 was a snapshot program, we could not guarantee exactly which targets
would be observed for this comparison. In the end, targets SMP~1 (N~1), SMP~6
(N~6), and SMP~22 (N~67) were re-observed with STIS. The first two targets 
were observed with PC1 earlier, while SMP~22 had been observed with FOC.

SMP~1 and SMP~6 are small nebulae with very uncomplicated structures. The round (E?) morphology
of these two targets was easily seen in the PC1 data set \citep{vasea98}. 
The early photometric radius measured for these two PNs were 0.12\arcsec~and 0.152\arcsec,
respectively for SMP~1 and SMP~6 was close to our measurements of 0.15\arcsec~and 0.19\arcsec. 

The situation of the bipolar planetary nebula SMP~22 is very different. 
The complete nebular morphology that we see in most emission lines of the STIS
spectra, but in particular in the \nii lines, is not as evident in the FOC image
\citep{sta99}, Figure 5. Furthermore, we measure a photometric
radius of 0.4\arcsec, while the radius from the FOC image is almost three times larger. 
We reanalyzed  these measurements, and noted
that a radius of 0.4\arcsec~ already included about 75$\%$ of the total flux, but 
to encircle the 85$\%$ of the flux required 1.37\arcsec~ radius.
From the comparison of the old and new SMP~22 images we conclude that the main morphological
features of MC PNs are reliable from the pre-COSTAR images, even if
the detailed morphology was not resolved. On the other hand, the measurements of
the photometric radii are not reliable.

\subsection{Statistics of morphological types }

From our analysis we can see that SMC PNs are nicely classified using the same 
morphological scheme as the LMC 
and Galactic PNs. In order to increase the sample size for statistical
purposes, we include in the present sample three additional SMC PNs described and 
classified in \citet{sta99}. After eliminating unresolved objects
and repeats, 
we have at our disposal a sample of 30 SMC PNs whose morphology is well determined.
This SMC PN sample constitute nearly 50 percent of all
known SMC PNs, thus we consider fairly representative at least of the bright PNs. 
In Table 4 we give the statistics of
PN morphology for the SMC sample as compared to the LMC and Galactic samples
from \citet{sha01}.

The LMC and SMC samples in columns 2 and 3 of Table 4, have been selected in similar ways,
and they have similar observational biases (but none of the
extreme selection biases that affect Galactic PN samples toward the Galactic
plane). 
While the fraction of round PNs remains more or less the same in the two samples,
the fractions of E and BC in the SMC are respectively twice and one-half
those of the LMC. The overall frequency of asymmetric PN in the SMC is only
sixty- percent that of the LMC. This remarkable results strongly suggests
that the
difference between the Magellanic Clouds is reflected in their PN
populations: something in the environment of the SMC may not
not be favorable to the formation of bipolar and bipolar core PNs.
Alternately, this may be the result of the low metallicity of the SMC,
indicating that the production of the higher
mass progenitors of PNs in the SMC has long subsided, and that the
low percentage of asymmetric PNs in the SMC is due to the lack of recent star
formation episodes. This issue will be better framed once we have complete
abundance analysis for the PNs analyzed in this paper.

\subsection{Surface brightness--radius relation}

We plot the surface brightness versus radius relation 
for SMC (Fig. 8) and SMC and LMC PNs together, using
the LMC data from \citet{sta02} (Fig.~9). In Figures 8 and 9
the PNs 
are coded for their morphological types. We exclude from these
plots round and elliptical PNs (both SMC and LMC) 
with inner (and outer) structure, 
to avoid possible misclassifications (see above). These figures confirm for SMC
PNs the trend that we found for LMC PNs: surface brightness in the
light of  \oiii $\lambda$5007, \ha, \hb, \oi $\lambda$6300, \nii $\lambda$6584, and \siii $\lambda$6312 evolves differently for 
the different morphological types. The surface brightness to
radius relation is very tight in all spectral lines, with the 
exception of the \nii emission line, where a larger spread is present.
From the lower left panels of Figures 8 and 9 we see that the
\nii spread is more extreme for bipolar PNs. A possible factor is the
higher but more varied nitrogen abundances of bipolar and BC PNs.

\subsection{The \oiii 5007/\hb distribution}

In Figure 10 we plot a histogram of the ratio of reddening-corrected 
fluxes of the \oiii $\lambda$5007 and \hb lines (hereafter ``\oiii/\hb'') 
for the PNs of the SMC and the LMC.  The median of the 
SMC distribution is a factor of two lower than for the corresponding LMC 
distribution.  Specifically, our STIS data yield galaxy averages 
$<$\oiii/\hb$>_{\rm SMC}$ = 5.7 $\pm$ 2.5 and $<$\oiii/\hb$>_{\rm LMC}$ = 9.4 $\pm$ 3.1.  

To the best of our knowledge this result is free of object selection 
biases since both sets of targets were chosen in much the same way.  The 
objects with very low \oiii/\hb ratios are very low-ionization 
objects whose central stars are presumably too cool to form much 
O$\rm^{++}$.  However, the \oiii/\hb ratio tends to reflect that of 
the PNs with the brightest \oiii and \hb lines.  In general we tend to favor 
targets with hottest 
central stars: T$_{eff} \ge 50,000$ K.

The low median value of \oiii/\hb in the SMC has been noted before,
from ground-based measurements \citep{webster}.  
However, the STIS images allow us to distinguish between small, bright HII 
regions and PNs.  Using the ground-based data from \citet{srmc} 
we derive substantially the same result: 
$<$\oiii/\hb$>_{\rm SMC}$ = 4.0 $\pm$ 2.8 and $<$\oiii/\hb$>_{\rm LMC}$ = 9.2 
$\pm$ 4.2.  For reference, the \oiii/\hb ratio for Galactic PNs is 
of the order of 15.

The \oiii/\hb emissivity ratio is physically scaled linearly with 
the O/H abundance and the fractional ionization of O$\rm^{++}$. Also it 
depends exponentially on the local electron excitation temperature, 
T$_e$(O$\rm^{++}$) since electron collisions on the high-energy tail of 
the free energy distribution excite the transition.  Of course, 
T$_e$(O$\rm^{++})$ depends on O/H and O$\rm^{++}$/O as well.  So 
interpreting the differences between the \oiii/\hb ratios of the 
SMC and the LMC is best done using ionization models.

We used the ionization model Cloudy \citep{cloudy} to understand the 
systematic trends in the behavior of \oiii/\hb for the SMC, LMC and 
the Galaxy.  We adopted a gas density of 1000 cm$^{-3}$ and 
standard chemical abundances for the three environments.  Galactic 
abundances for the PNs of the Galaxy are those adopted in the Paris 
meeting (1985; see Table 7 in the Hazy manual, Ferland 1996).  
We used LMC average abundances of PNs quoted in \citet{sta00}
and SMC average abundances from \citet{srmc} except for the C/H ratio
which comes from \citet{ld96}, where we had
selected from their Table 3 only the low-error data. 
The stellar ionizing spectrum 
is assumed to be a blackbody with temperatures and luminosities from the 
H-burning evolutionary tracks for the appropriate galaxian population
by \citet{vw}.  Our model parameters are summarized in 
Table 5.

The predictions of \oiii/\hb from Cloudy models are shown in Figure
11 for the SMC, LMC, and Galactic PNs.  The outcomes of the Cloudy models 
are rather insensitive to the adopted stellar properties and the assumed 
density (e.g., Fig. 12).  But they are very sensitive to the adopted  
abundance ratios, especially the oxygen abundance.  
The model values of \oiii/\hb are shown as a 
function of stellar temperature over the range of temperatures encountered 
by a star of mass = 2 \sm, as it evolves from the AGB tip
to its maximum post-AGB temperature.  At stellar temperatures in 
excess of $\sim10^5$ K the \oiii/\hb ratio is in rough agreement 
with the present observations (see also \citet{gda,gdb}). It is worth noting that
the Galactic models presented here are not to be compared directly with the known 
observed \oiii/\hb distribution, without accounting for selection effects that
hamper Galactic PN statistics.
We should clarify that the results of Figure 11 are valid for the
input abundances. If, for example, we chose to use as input the average oxygen
abundance for SMC ONs from \citet{ld96} instead of that of \citet{srmc},
the \oiii~over \hb~ emission line would be as high as 7.2. 

The cooling processes that determine T$_e$(O$\rm^{++})$ in the SMC, LMC 
and Galactic PNs are noteworthy.  In the Galaxy the primary coolants of 
PNs with hot central stars are the optical forbidden lines of \oiii
$\lambda$5007 and other lines of O$\rm^+$ and O$\rm^{++}$.  However, 
in environments in which O/H is as low as in the SMC, the primary 
coolants may become ultraviolet intercombination lines of 
C$\rm^+$ and C$\rm^{++}$.  It will be interesting to confirm these 
predictions with future UV observations.

Before reaching conclusions and further speculation, let us explore the weight of
our assumptions. 
One is that these models have constant hydrogen 
density, which is arbitrarily assumed equal to 1000 cm$^{-3}$. By running Cloudy
for a constant density model, the outer radius is determined by the Stroemgen sphere. 
This makes our models larger as stellar temperature increases, reaching a 
maximum and then decline with the \oiii $\lambda$5007/\hb intensity ratio. But in order to compare the radii
of our models with our data we should make a model for each nebula with the correct
radius (i.e., assuming a density profile that reproduces that PN, for example, as 
determined in hydrodynamic calculations). This 
will be done in detail in a future paper. For now, let us examine what a difference
in the average hydrogen density will make in the \oiii 5007 flux ratio. In Figure 12
we plot the \oiii/\hb ratio against the oxygen abundance for the early and the 
hottest models in the SMC (triangles), LMC (squares), and the Galaxy (circles). 
The early models correspond to models 1, 6, and 11 of Table 5, with log N$_{\rm H}$=3 cm$^{-3}$. 
The hot models have been calculated for log N$_{\rm H}$=2.5, 3, and 3.5 cm$^{-3}$.
We see that varying the density does not affect very much the studied intensity ratio,
especially for the low oxygen models. The different hot SMC models in Figure 12
are all within $\delta$ I(5007)/I(\hb)=1, and their outer radii are 0.11, 0.25, and 0.55 parsecs
for log N$_{\rm H}$=3.5, 3, and 2.5 sm$^{-3}$ respectively. 

All models described in this paper have filling factor ($\epsilon$) equal to unity. Changing the value
of the filling factor will also change the outer radius of the nebular models. For
example, the hot Galactic model (with log N$_{\rm H}$=3) with $\epsilon$=0.5 has
an outer radius about 1.3 times larger than the model with $\epsilon$=1. We find that 
the \oiii $\lambda$5007/\hb intensity ratio with the different radius and filling factor
is 95$\%$ the ratio
in the unity filling factor model, thus the filling factor assumption is 
not very important the discussion of the \oiii 5007 line intensity with respect to
\hb. 

Oxygen is usually the major coolant in the oxygen abundant PNs.
But what other coolants play a role? In Figure 13 we show the intensity of the line relative to \hb
for the major coolants in the SMC, LMC, and Galactic PNs, versus the oxygen abundance.
The intensities are from the hottest models for each galaxian mix.
Cloudy predicts that although the 5007 \AA~ line is the major coolant for the Galactic and LMC
PNs, in the SMC PNs the \civ $\lambda$1548 line is the major coolant. Other
carbon lines are also important coolant in the SMC PNs, while the \oiii $\lambda$4959 and \oii
$\lambda$3727 lines are significant mostly for the Galactic models. Clearly, more detailed models and observations
of \civ $\lambda$1909 and \civ $\lambda$1550 are essential to confirm these predictions, but 
the interpretation is plausible and consistent with the available data.

In \citet{sta02} we used the I\oiii (5007+4959) / \ha ratio to trace the nebular excitation,
and by inference the stellar temperature, of the LMC PNs. We can not perform the same analysis here,
since it is clear from Figure 11 that the \oiii/\hb  ratio versus temperature relation is
very non-monotonic. 

\subsection{Electron densities and ionized masses}
 
Table 6 lists nebular densities determined from strong \sii $\lambda\lambda$6716,6731
lines for eight SMC PNs (this paper) and twelve LMC PNs \citep{sta02}. We assumed T$_{\rm e}$=10$^4$ K
for these estimates. However, the derived densities are extremely insensitive to T$_{\rm e}$
\citep{ost89}. We find that $<{\rm N}_{\rm e}>_{\rm SMC}$=3.45, whereas $<{\rm N}_{\rm e}>_{\rm LMC}=$3.28, a 
factor 1.5 different. Given the small sample size, a factor of 1.5 is not likely to be
very significant. Furthermore, there is no discernible trend of N$_{\rm e}$ with 
morphological type in SMC or LMC PNs. Moreover, nebulae with bright \sii lines are
strongly biased to those of generally high fluxes or surface brightness, or low ionization.
These selection biases render trends in N$_{\rm e}$ of limited significance.

In addition we estimate masses of PNs in the LMC and SMC using the method of \citet{bs94}.
The average ionized mass of the eight SMC PNs, 0.3 \sm, is slightly larger than that of
LMC PNs, 0.2 \sm. Given the uncertainty in the data, and the small data sample, we do not believe that
the mass discrepancy between SMC and LMC PNs is significant. Also, we fail to see any obvious trend between 
nebular mass and morphological type.

\section{Summary and conclusions}

A sample of 27 SMC PNs have been observed in imaging a slitless mode with {\it HST}/STIS
to examine their morphology, shape, and fluxes, and to study their evolution.
This morphological sample is the first sizable set of SMC PNs, and represents
almost half of the known SMC PNs. 
The images and spectra have the same high quality and resolution as our
LMC sample. We present the broad images and monochromatic images in the major
emission lines, and determine that morphology is easily recognized in most 
emission lines. We find that the ratio of symmetric to asymmetric
PNs is remarkably different in the SMC and the 
LMC. Specifically, bipolar PNs are much rarer in the SMC
than the LMC (or the Galaxy). This new result has significant implications for
the relation between stellar population and PN morphology. 
It is well known from Galactic and LMC PN studies that PN morphology correlates
with the mass of the progenitor stars. In particular, bipolar PNs evolve from relatively massive
($\ge 1.5$ \sm) progenitors. Thus the low incidence of bipolar PNs in the SMC probably reflects
the low formation rate of these stars in the past $\ge$5 Gyr. This is in accord with
other studies of star formation rates in the LMC and the SMC. Alternately, or in addition,
the low metallicity of the SMC may inhibit wind collimation somehow.

We also present the measurement of the optical line intensities.
We find that the surface brightness declines with radii in most emission lines,
adding value to the possible calibration of the surface brightness- nebular radius 
correlation seen in the LMC. 
Ionized masses and electron densities were
calculated where the \sii doublet intensity was available. The resulting ionized
masses do not seem to have a relation to the morphology, as already
found in the LMC PNs.

The \oiii/\hb ratio of Magellanic Cloud PNs has been studied in detail. The factor
of two difference
between the LMC and SMC PNs were modeled with Cloudy.  We find that the 
cooling models strongly depend on the oxygen content, and that
the brightest parts of the \oiii 5007/\hb luminosity function shifts 
between SMC and LMC PNs. We also find a relation between the brightest 
\oiii 5007/\hb PNs and their morphology, that is different for each Magellanic
Cloud population. 

\acknowledgments 

We thank George Jacoby and Orsola de Marco for discussing the faint target selection, 
Orsola de Marco for her help in completing the Phase II of the {\it HST} program,
and Massimo Stiavelli for his suggestions about Cloudy models.
This work was supported by NASA through grant GO-08663.01-A from Space
Telescope Science Institute, which is operated by the Association of 
Universities for Research in Astronomy, Incorporated, under NASA contract
NAS -26555. 

\clearpage

\figcaption{STIS images of the LMC PNs J~4, J~18, J~23, and
J~27 with a logarithm intensity scale. From left to right, we show for each PN the broad band image, the 
\nii -- \ha section of the G750M spectrum, and the \oiii 5007 image of the
G430M spectrum, when available.}


\figcaption{Same as in Figure 1, but for MA~1682, MA~1762, MG~8, and MG~13.}

\figcaption{Same as in Figure 1, but for SMP~1, SMP~6, SMP~8, SMP~9, and
  SMP~11.}

\figcaption{Same as in Figure 1, but for SMP~12, SMP~13, SMP~14, SMP~17, and
  SMP~18.}

\figcaption{Same as in Figure 1, but for SMP~19, SMP~20, SMP~22, SMP~23, and
  SMP~24.}

\figcaption{Same as in Figure 1, but for SMP~25, SMP~26, SMP~27, SP~34.}

\begin{figure}
\epsscale{0.6}
\plotone{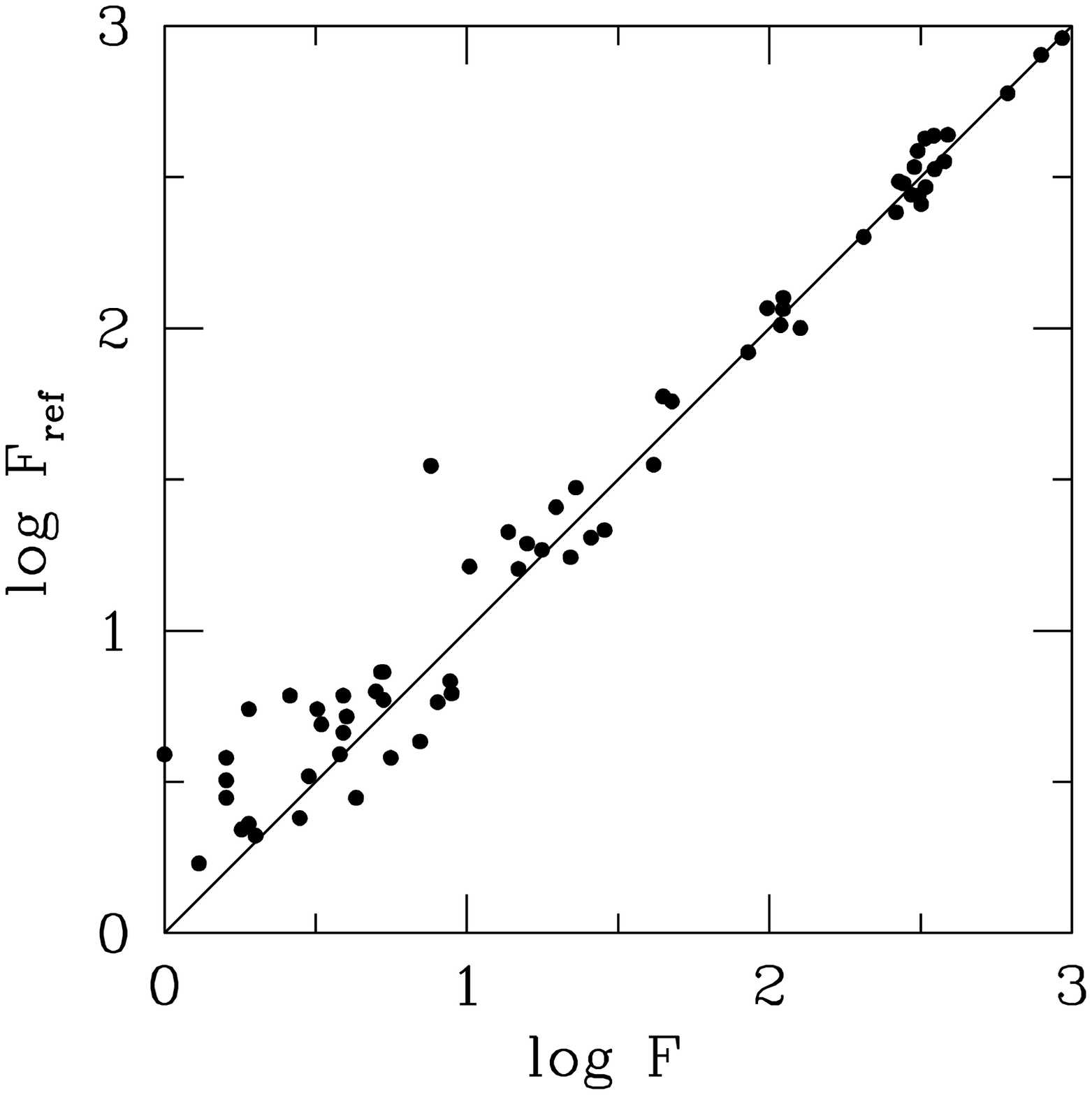}
\caption[]{Comparison of the measured line intensity ratios with those in the literature,
in logarithmic scale.
\label{f7.eps}}
\end{figure}

\begin{figure}
\epsscale{1}
\plotone{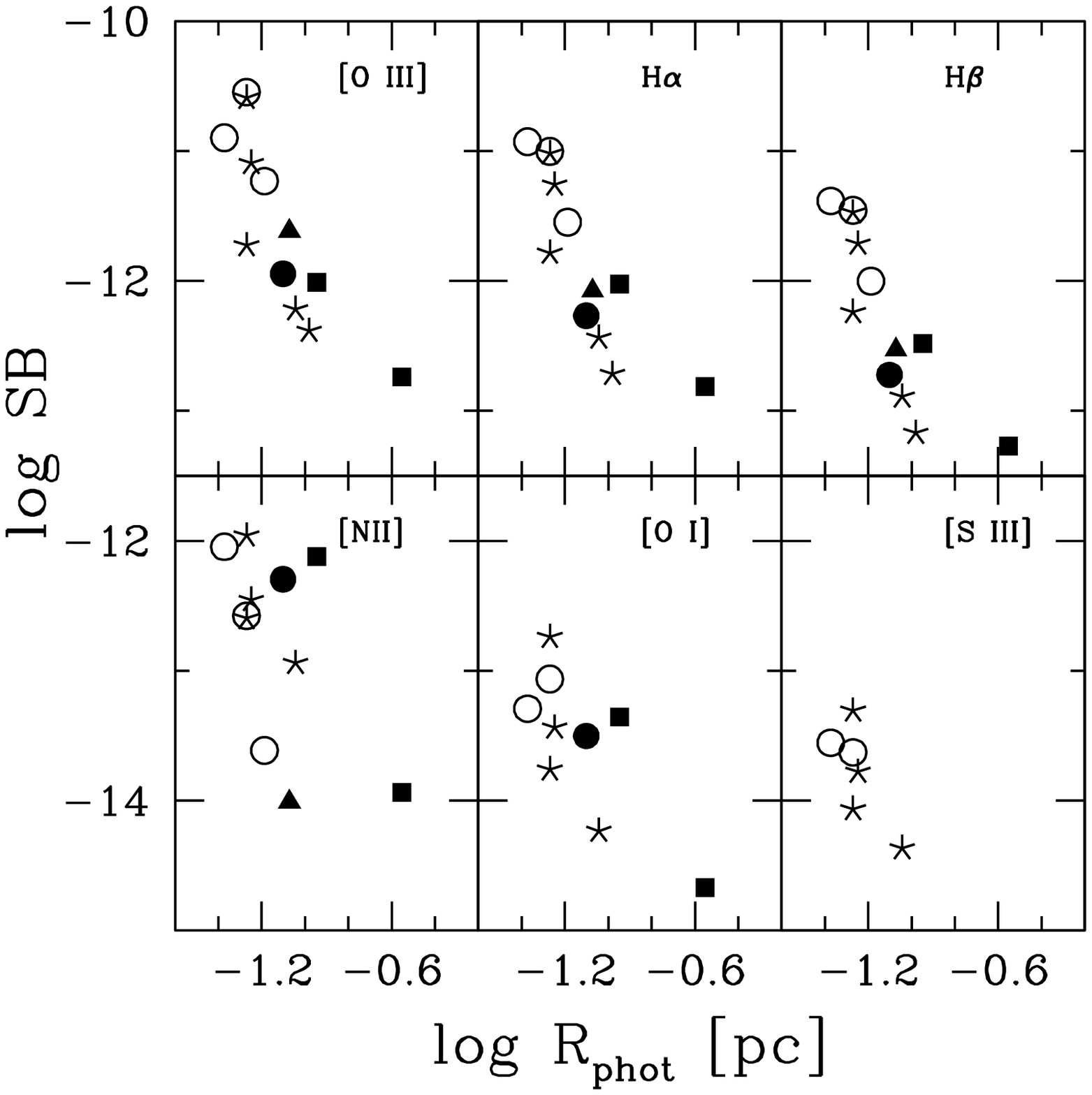}
\caption[ ]{Surface brightness decline for the multiwavelength images of the PNs in our 
STIS survey. Emission lines in which the SB is derived are indicated in the panels. 
Symbols indicate morphological types: open circles=round, filled circles=pointsymmetric,
stars=elliptical, filled triangles=bipolar core, filled squares=bipolar (and 
quadrupolar) planetary nebulae. The photometric radii are measured from the 
\oiii $\lambda$5007 images, where available.
\label{f8.eps}}
\end{figure}

\begin{figure}
\epsscale{1}
\plotone{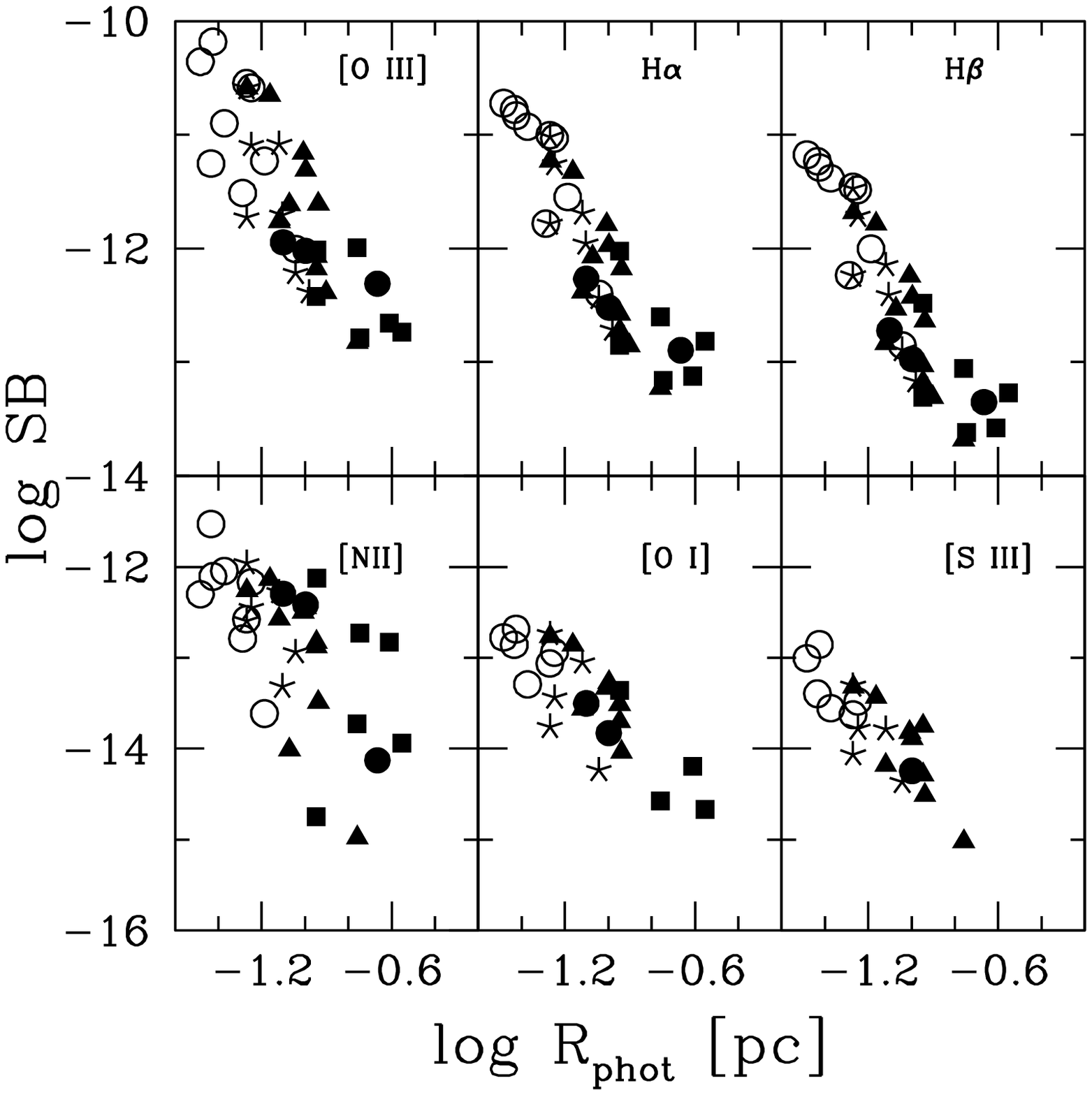}
\caption[]{Same as in Figure 8, but for SMC and LMC PNs together (see text).
\label{f9.eps}}
\end{figure}

\begin{figure}
\epsscale{0.5}
\plotone{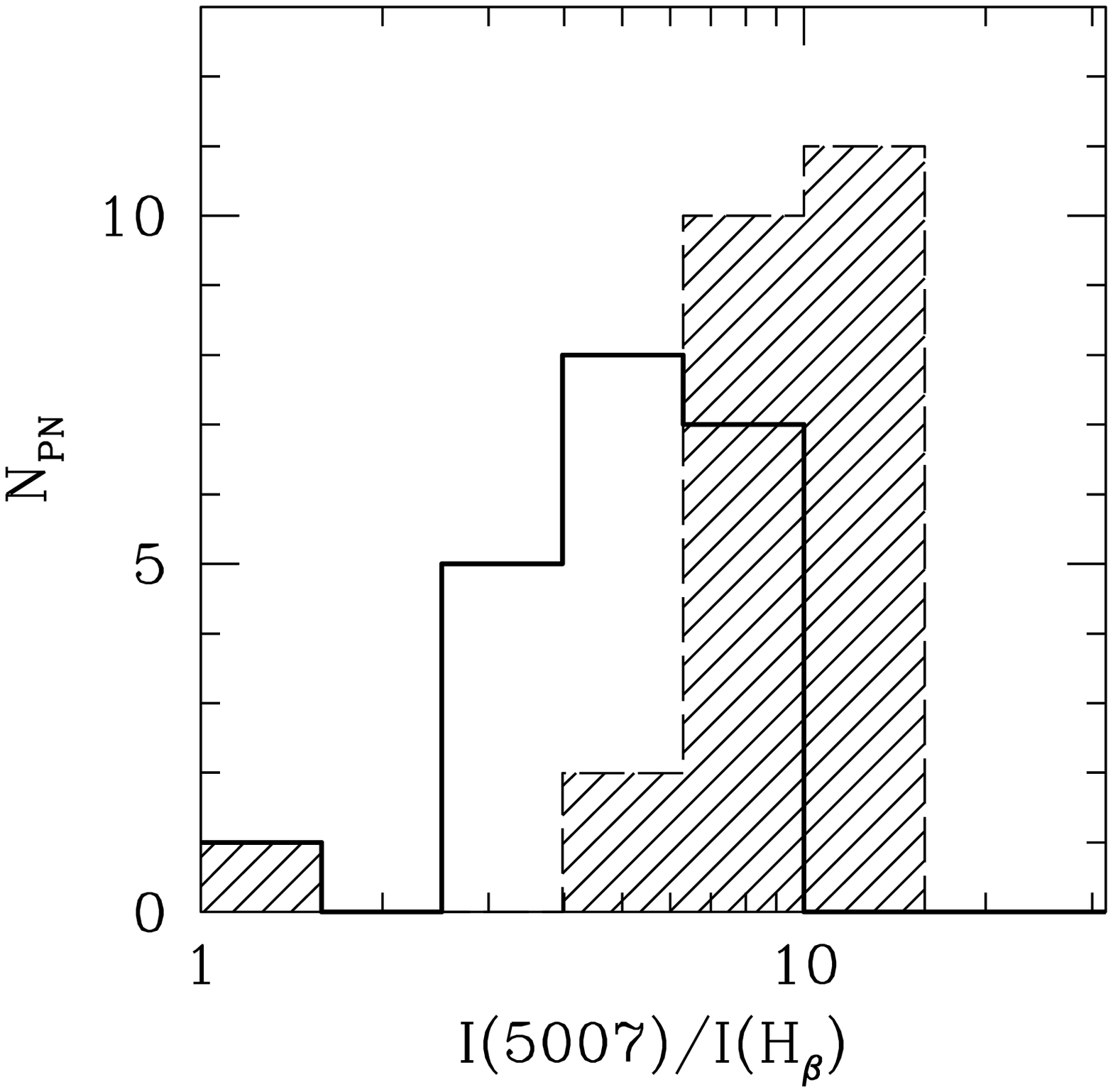}
\caption[]{Distribution of the \oiii $\lambda$5007 over \hb intensity ratios
in the SMC (thick histogram) and LMC (shaded histogram) PNs.
\label{f10.eps}}
\end{figure}

\begin{figure}
\epsscale{0.5}
\plotone{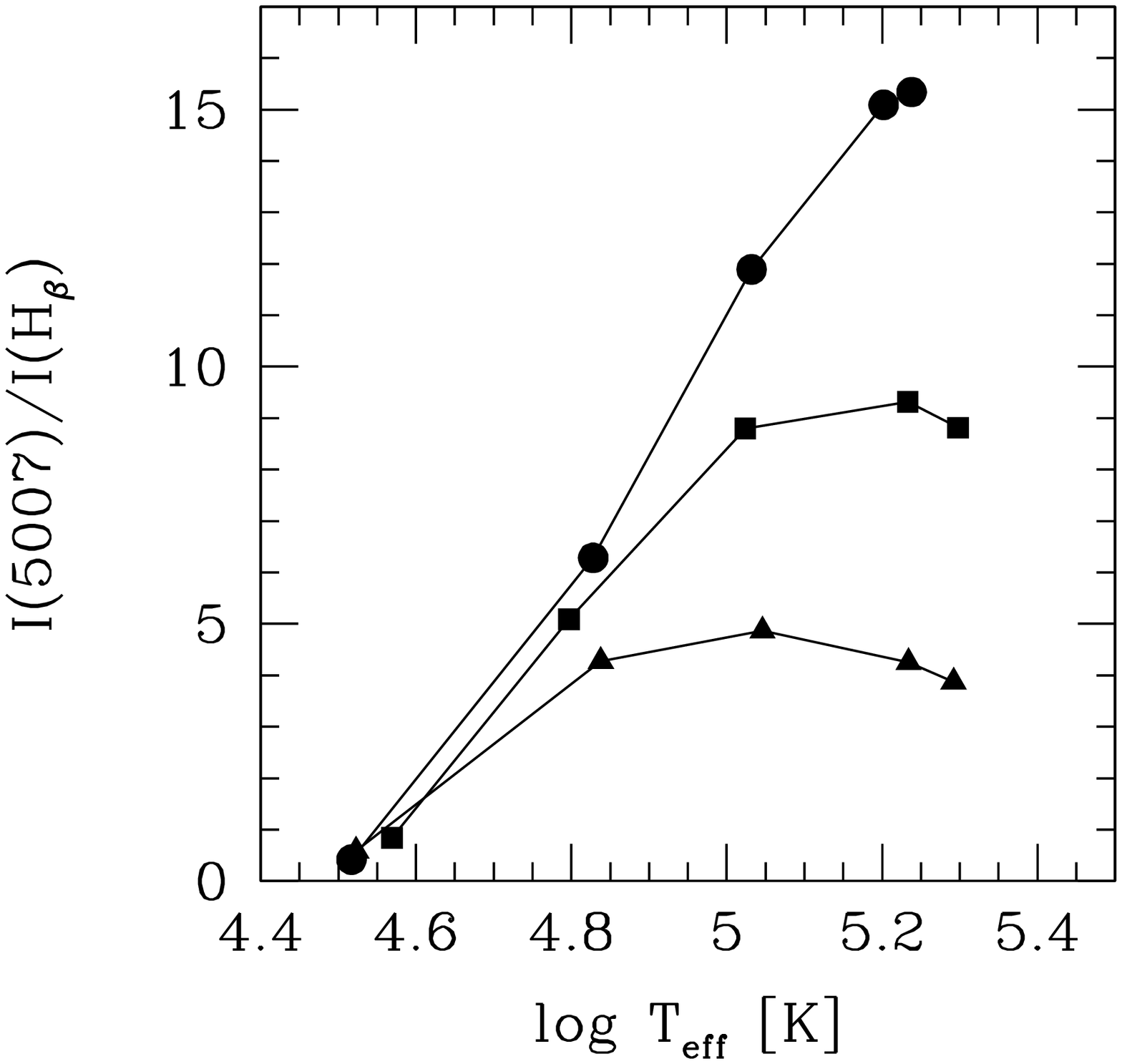}
\caption[]{The \oiii $\lambda$5007 over \hb intensity ratios for SMC (triangles),
LMC (squares) and Galactic (circles) models. Temperatures reflect the evolution from the AGB to
the maximum central star temperature in the evolutionary tracks of SMC, LMCm and Galactic
post-AGB stars.
\label{f11.eps}}
\end{figure}

\begin{figure}
\epsscale{0.5}
\plotone{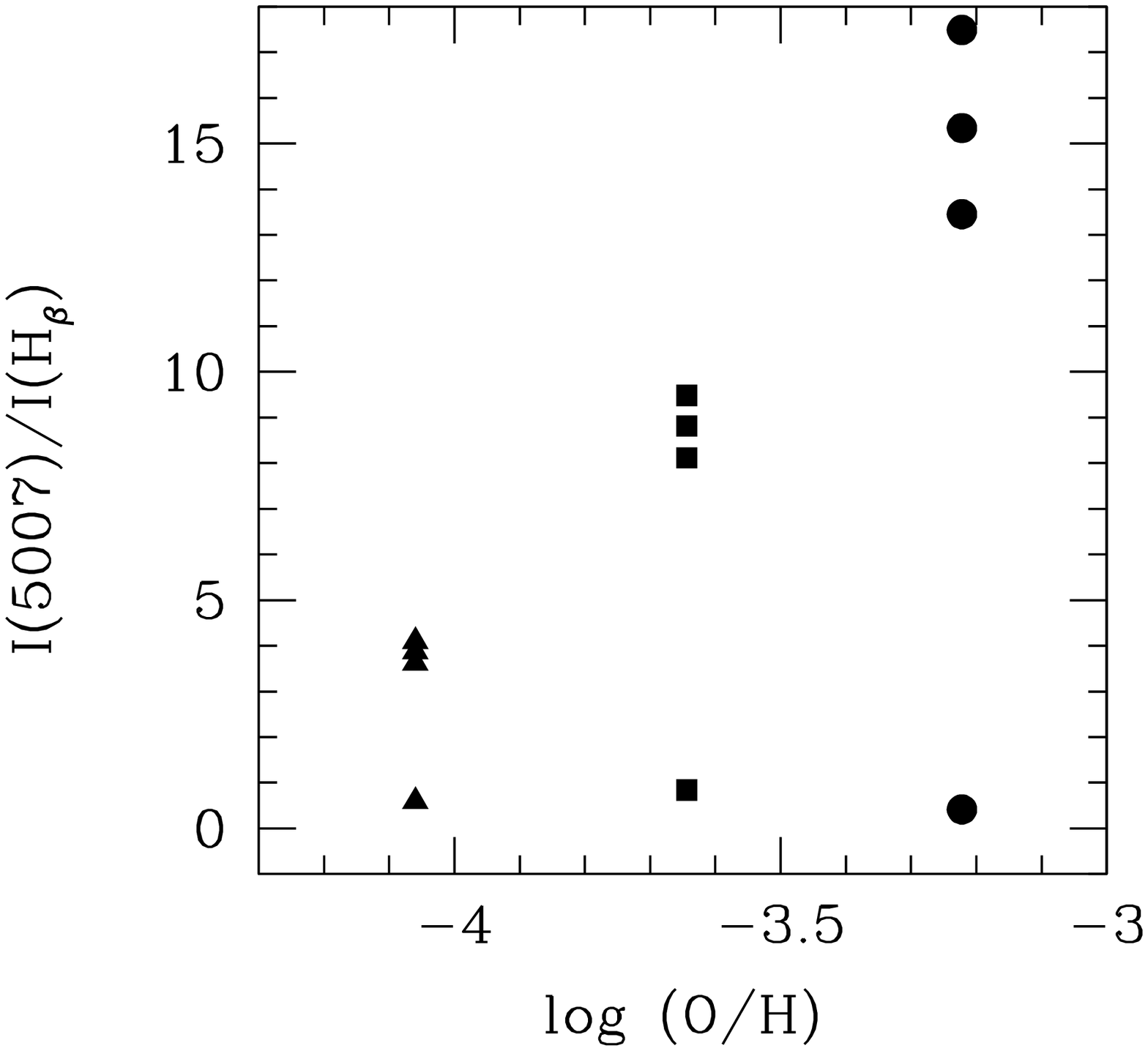}
\caption[]{The \oiii $\lambda$5007 over \hb intensity ratios versus the
oxygen abundances for SMC, LMC, and Galactic PNs. Symbols as in Figure 11.
PN models have constant density.
We plot one cool model, and a group of hot models, for each galaxy. The three hot models
represent three mean densities, see text.
\label{f1.eps}}
\end{figure}

\begin{figure}
\epsscale{0.5}
\plotone{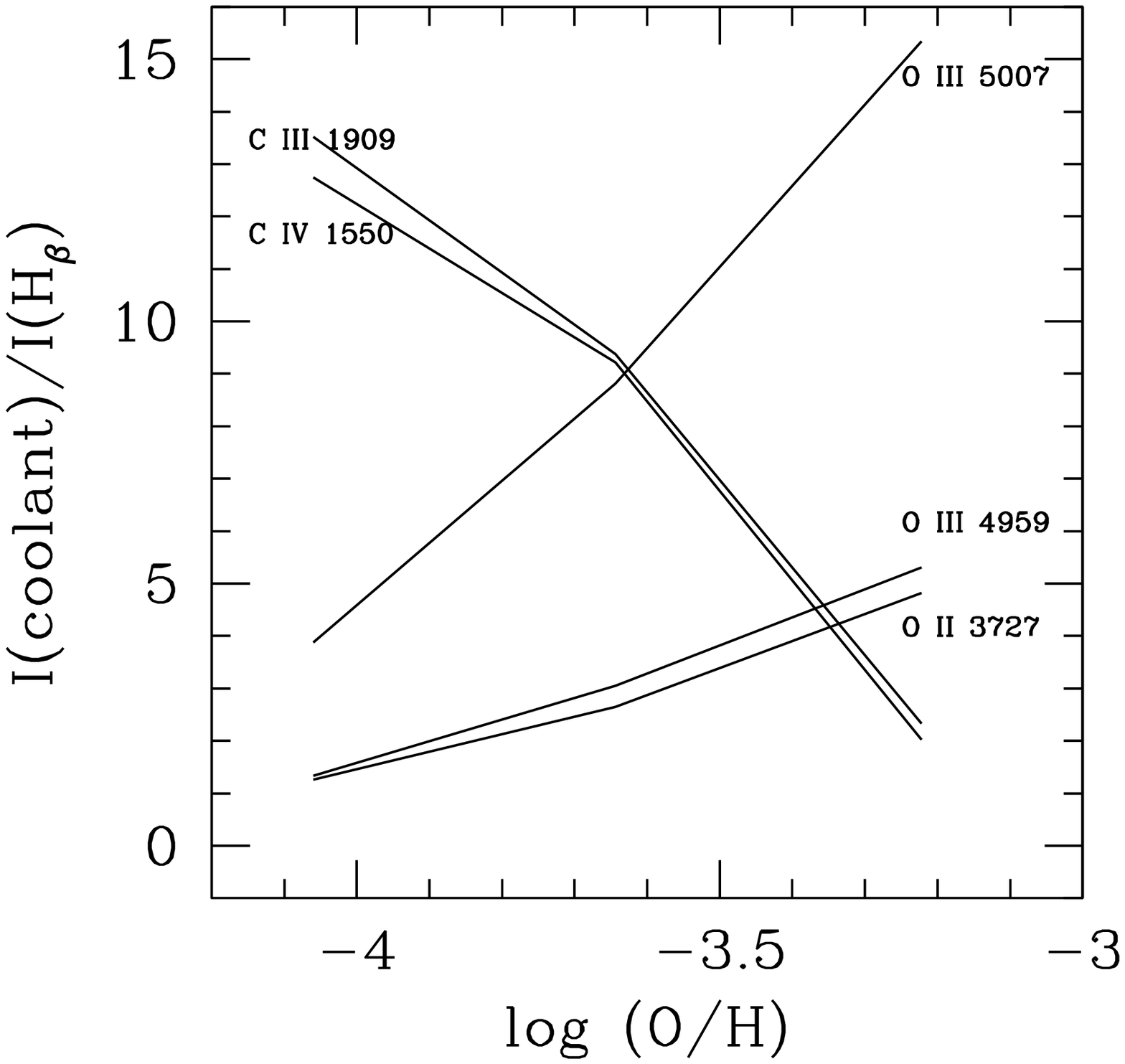}
\caption[]{Intensity ratios of the major PN coolants over \hb, versus oxygen
abundances. The coolant is indicated in the right hand side of the plot\label{f13.eps}}
\end{figure}

\clearpage

\begin{deluxetable}{lcllrc}

\tabletypesize{\scriptsize}
\tablewidth{0pt}
\tablecaption {Observing Log for SMC Planetary Nebulae \label{ObsLog}}
\tablehead {
\colhead {} & \colhead {} & \colhead {} & 
\colhead {} & \colhead {T$_{Exp}$} & \colhead {} \\
\colhead {Nebula} & \colhead {Date} & \colhead {Data Set} & 
\colhead {Disperser} & \colhead {(s)} & \colhead {N$_{Exp}$} 
}
\startdata  

J 4       & 2000 Aug 18  & O65S03030  & G430M   &  840  & 2 \\
          &              & O65S03020  & G750M   &  890  & 2 \\
          &              & O65S03010  & MIRVIS  &  300  & 2 \\
J 18      & 2000 Oct 3   & O65S08020  & G750M   & 2100  & 2 \\
          &              & O65S08010  & MIRVIS  &  300  & 2 \\
J 23      & 2000 Aug 8   & O65S12020  & G750M   & 2100  & 2 \\
          &              & O65S12010  & MIRVIS  &  300  & 2 \\
J 27      & 2000 Oct 10  & O65S16020  & G750M   & 2100  & 2 \\
          &              & O65S16010  & MIRVIS  &  300  & 2 \\
MA 1682   & 2000 Nov 5   & O65S24020  & G750M   & 2100  & 2 \\
          &              & O65S24010  & MIRVIS  &  300  & 2 \\
MA 1762   & 2000 Aug 19  & O65S25020  & G750M   & 2100  & 2 \\
          &              & O65S25010  & MIRVIS  &  300  & 2 \\
MA 1796   & 2000 Sep 27  & O65S26040  & G430M   &  200  & 2 \\
          &              & O65S26020  & G750M   &  100  & 2 \\
          &              & O65S26030  & G750M   &  400  & 2 \\
          &              & O65S26010  & MIRVIS  &  120  & 2 \\
          &              & O65S26KGQ  & MIRVIS  &   15  & 2 \\
MG 2      & 2000 Nov 7   & O65S28030  & G430M   &  800  & 2 \\
          &              & O65S28020  & G750M   & 1300  & 2 \\
          &              & O65S28010  & MIRVIS  &  300  & 2 \\
MG 8      & 2000 Nov 7   & O65S34030  & G430M   & 1200  & 2 \\
          &              & O65S34020  & G750M   &  620  & 2 \\
          &              & O65S34010  & MIRVIS  &  120  & 2 \\
MG 13     & 2000 Sep 28  & O65S39020  & G750M   & 2100  & 2 \\
          &              & O65S39010  & MIRVIS  &  300  & 2 \\
SMP 1     & 2001 Feb 6   & O65S40030  & G430M   &  360  & 2 \\
          &              & O65S40020  & G750M   &  170  & 2 \\
          &              & O65S40010  & MIRVIS  &  120  & 2 \\
SMP 4     & 2001 Jan 9   & O65S41030  & G430M   &  460  & 2 \\
          &              & O65S41020  & G750M   &  230  & 2 \\
          &              & O65S41010  & MIRVIS  &  120  & 2 \\
SMP 6     & 2001 Mar 9   & O65S42030  & G430M   &  280  & 2 \\
          &              & O65S42020  & G750M   &  190  & 2 \\
          &              & O65S42010  & MIRVIS  &  120  & 2 \\
SMP 8     & 2001 Jan 23  & O65S43030  & G430M   &  320  & 2 \\
          &              & O65S43020  & G750M   &  160  & 2 \\
          &              & O65S43010  & MIRVIS  &  120  & 2 \\
          &              & O65S43GKQ  & MIRVIS  &   15  & 1 \\
SMP 9\tablenotemark{a}    & 2000 Nov 20  & O65S02030  & G430M   & 1100  & 2 \\
          &              & O65S02020  & G750M   &  790  & 2 \\
          &              & O65S02010  & MIRVIS  &  300  & 2 \\
SMP 11\tablenotemark{a}    & 2000 Sep 6   & O65S05040  & G430M   &  200  & 2 \\
          &              & O65S05020  & G750M   &   60  & 2 \\
          &              & O65S05030  & G750M   &  300  & 2 \\
          &              & O65S05010  & MIRVIS  &  120  & 2 \\
          &              & O65S05LBQ  & MIRVIS  &   15  & 1 \\
SMP 12    & 2000 Oct 14  & O65S44030  & G430M   &  650  & 2 \\
          &              & O65S44020  & G750M   &  810  & 2 \\
          &              & O65S44010  & MIRVIS  &  300  & 2 \\
SMP 13\tablenotemark{a}     & 2000 Nov 20  & O65S06040  & G430M   &   52  & 2 \\
          &              & O65S06020  & G750M   &   36  & 2 \\
          &              & O65S06030  & G750M   &  180  & 2 \\
          &              & O65S06010  & MIRVIS  &  120  & 2 \\
          &              & O65S06CHQ  & MIRVIS  &   15  & 1 \\
SMP 14    & 2000 Sep 17  & O65S45030  & G430M   &  540  & 2 \\
          &              & O65S45020  & G750M   &  290  & 2 \\
          &              & O65S45010  & MIRVIS  &  120  & 2 \\
SMP 17    & 2001 Jan 16  & O65S47040  & G430M   &  160  & 2 \\
          &              & O65S47020  & G750M   &  100  & 2 \\
          &              & O65S47030  & G750M   &  100  & 2 \\
          &              & O65S47010  & MIRVIS  &  120  & 2 \\
          &              & O65S47FXQ  & MIRVIS  &   15  & 1 \\
SMP 18\tablenotemark{a}  & 2001 May 16  & O65S09040  & G430M   &  160  & 2 \\
          &              & O65S09020  & G750M   &   42  & 2 \\
          &              & O65S09030  & G750M   &  210  & 2 \\
          &              & O65S09010  & MIRVIS  &  120  & 2 \\
          &              & O65S09JIQ  & MIRVIS  &   15  & 1 \\
SMP 19\tablenotemark{a}    & 2000 Oct 14  & O65S10030  & G430M   &  330  & 2 \\
          &              & O65S10020  & G750M   &  260  & 2 \\
          &              & O65S10010  & MIRVIS  &  120  & 2 \\
SMP 20    & 2001 Jan 21  & O65S48040  & G430M   &  160  & 2 \\
          &              & O65S48020  & G750M   &   80  & 2 \\
          &              & O65S48030  & G750M   &  320  & 2 \\
          &              & O65S48010  & MIRVIS  &  120  & 2 \\
          &              & O65S48YXQ  & MIRVIS  &   15  & 1 \\
SMP 22    & 2000 Oct 14  & O65S49030  & G430M   &  840  & 2 \\
          &              & O65S49020  & G750M   &  200  & 2 \\
          &              & O65S49010  & MIRVIS  &  120  & 2 \\
SMP 23\tablenotemark{a}    & 2000 Oct 14  & O65S15030  & G430M   &  230  & 2 \\
          &              & O65S15020  & G750M   &  380  & 2 \\
          &              & O65S15010  & MIRVIS  &  120  & 2 \\
SMP 24    & 2001 Mar 9   & O65S50030  & G430M   &  500  & 2 \\
          &              & O65S50020  & G750M   &  150  & 2 \\
          &              & O65S50010  & MIRVIS  &  120  & 2 \\
SMP 25    & 2001 Mar 5   & O65S51030  & G430M   &  920  & 2 \\
          &              & O65S51020  & G750M   &  460  & 2 \\
          &              & O65S51010  & MIRVIS  &  120  & 2 \\
SMP 26    & 2000 Sep 28  & O65S52030  & G430M   & 1200  & 2 \\
          &              & O65S52020  & G750M   &  900  & 2 \\
          &              & O65S52010  & MIRVIS  &  300  & 2 \\
SMP 27    & 2001 Jan 19  & O65S53040  & G430M   &  160  & 2 \\
          &              & O65S53020  & G750M   &   80  & 2 \\
          &              & O65S53030  & G750M   &  360  & 2 \\
          &              & O65S53010  & MIRVIS  &  120  & 2 \\
          &              & O65S53QBQ  & MIRVIS  &   15  & 1 \\
SP 34     & 2000 Oct 14  & O65S55030  & G430M   & 1000  & 2 \\
          &              & O65S55020  & G750M   &  500  & 2 \\
          &              & O65S55010  & MIRVIS  &  300  & 2 \\
          \enddata  

\tablenotetext{a}{SMP~9=J~3; SMP~11=J~8; SMP~13=J~11; SMP~18=J~19; SMP~19=J~20;SMP~23=J~26}
\
\end{deluxetable}

\clearpage
\newpage
\begin{deluxetable}{lrrrrrrrrrrrrr}
\tabletypesize{\scriptsize}
\tablewidth{0pt}
\tablecaption {Relative Emission Line Intensities of SMC Planetary Nebulae \label{Flux}}
\tablehead {
\colhead {} & \colhead {$F$(H$\beta$)} & \colhead {} & \colhead {\oiii} & 
\colhead {\oiii} & \colhead {\oi} & \colhead {\siii} &
\colhead {\oi} & \colhead {\nii} & \colhead {H$\alpha$} &
\colhead {\nii} & \colhead {He I} & \colhead {\sii} &
\colhead {\sii} \\
\colhead {Nebula} & \colhead {(4861)} & \colhead {$c$} & \colhead {(4959)} & 
\colhead {(5007)} & \colhead {(6300)} & \colhead {6312} & \colhead {6363} & 
\colhead {(6548)} & \colhead {(6563)} & \colhead {6584} & \colhead {6678} & 
\colhead {(6716)} & \colhead {(6731)} \\
\colhead {(1)} & \colhead {(2)} & \colhead {(3)} & \colhead {(4)} & 
\colhead {(5)} & \colhead {(6)} & \colhead {(7)} & \colhead {(8)} & 
\colhead {(9)} & \colhead {(10)} & \colhead {(11)} & \colhead {(12)} & 
\colhead {(13)} & \colhead {(14)}  
}
\startdata  

J~4     &$-13.55$       &0.168  &159.1  &480.4  &5.1    &3.8    &2.5    &28.9   &326.3  &102.5  &1.9    &1.6    &3.0\\
J~18    &$-13.90$\tablenotemark{a}      &\nodata& \nodata& \nodata& 1.736& \nodata&  \nodata& 107.2&   100&    330.4&     \nodata&  4.2&       2.928\\
J~23    &$-13.19$\tablenotemark{a}      &\nodata& \nodata& \nodata& \nodata& \nodata& \nodata&  \nodata& 100& 13.9& \nodata& \nodata& \nodata\\
J~27    &$-14.05$\tablenotemark{a,b}    &\nodata& \nodata& \nodata&\nodata& \nodata& \nodata&  \nodata& 100& \nodata&\nodata& \nodata& \nodata\\
MA~1682& $-14.17$\tablenotemark{a}      &\nodata& \nodata& \nodata&\nodata& \nodata& \nodata&  \nodata& 100& 304.5& \nodata& \nodata\\
MA~1762& $-13.36$\tablenotemark{a}      &\nodata& \nodata& \nodata&  0.7829& \nodata& \nodata&\nodata& 100& \nodata&\nodata& \nodata& \nodata\\
MG~8    &$-13.27$       &0.133  &41.4   &126.5  &2.0    &1.6    &\nodata        &22.9   &317.2  &98.3   &4.7    &5.3    &22.0\\
MG~13   &$-13.10$\tablenotemark{a}      &\nodata& \nodata& \nodata&\nodata& \nodata&\nodata& \nodata& 100& \nodata& \nodata&\nodata&\nodata\\
SMP~1   &$-12.85$       &0.287  &85.0   &262.1  &2.0    &0.9    &0.7    &8.9    &359.3  &25.7   &4.3    &0.3    &0.7\\
SMP~6   &$-12.80$       &0.385  &267.7  &791.1  &7.0    &1.9    &1.6    &13.7   &388.6  &44.5   &5.3    &0.8    &1.6\\
SMP~8   &$-12.81$       &0.026  &198.7  &592.3  &\nodata        &\nodata        &\nodata        &1.3    &291.0  &2.5    &3.6    &\nodata        &\nodata\\
SMP~9   &$-13.46$       &0.070  &318.8  &959.4  &1.4    &1.6    &\nodata        &8.1    &301.4  &23.5   &\nodata        &9.7    &6.8\\
SMP~11  &$-13.13$       &0.352  &110.9  &351.4  &5.0    &\nodata        &\nodata        &10.0   &378.6  &28.5   &3.9    &3.3    &8.0\\
SMP~12  &$-13.59$       &0.056  &203.5  &613.3  &\nodata        &\nodata        &\nodata        &\nodata        &298.0  &\nodata        &3.5    &\nodata        &\nodata\\
SMP~13  &$-12.59$       &0.190  &277.3  &828.1  &2.8    &0.8    &1.0    &2.9    &332.0  &8.8    &3.9    &0.8    &1.3\\
SMP~14  &$-13.04$       &0.069  &311.1  &928.8  &2.6    &1.1    &1.0    &4.4    &301.2  &10.2   &1.0    &1.8    &3.0\\
SMP~17  &$-12.55$       &0.064  &294.0  &893.2  &2.8    &0.3    &1.2    &3.7    &300.0  &7.6    &4.0    &1.1    &0.5\\
SMP~18  &$-12.66$       &0.122  &105.0  &309.0  &1.3    &0.7    &0.4    &7.3    &314.5  &24.0   &3.6    &0.4    &0.7\\
SMP~19  &$-13.04$       &0.161  &282.4  &847.2  &4.0    &1.2    &0.9    &3.4    &324.5  &7.7    &2.4    &1.5    &2.2\\
SMP~20  &$-12.47$       &-0.019 &143.8  &434.9  &\nodata        &\nodata        &\nodata        &1.0    &280.8  &2.8    &3.9    &\nodata        &\nodata\\
SMP~22  &$-12.94$       &0.165  &99.1   &298.2  &14.8   &\nodata        &5.6    &87.7   &325.4  &259.6  &2.9    &9.8    &14.3\\
SMP~23  &$-13.18$       &0.101  &271.5  &828.1  &\nodata        &\nodata        &\nodata        &\nodata        &309.2  &3.6    &3.6    &\nodata        &\nodata\\
SMP~24  &$-12.66$       &0.047  &137.9  &420.1  &1.9    &0.9    &\nodata        &6.2    &295.9  &18.8   &4.0    &2.1    &2.5\\
SMP~25  &$-13.28$       &0.100  &108.8  &328.2  &3.2    &1.6    &\nodata        &15.8   &309.0  &47.6   &5.6    &1.3    &1.9\\
SMP~26  &$-13.58$       &0.253  &205.0  &613.0  &19.7   &\nodata        &5.2    &111.1  &349.4  &327.2  &2.0    &14.8   &17.7\\
SMP~27  &$-12.51$       &0.040  &188.8  &580.1  &1.4    &0.5    &0.4    &1.8    &294.2  &5.2    &3.8    &0.2    &0.6\\
SP~34   &$-13.67$       &0.163  &162.8  &488.4  &12.7   &\nodata        &22.4   &19.0   &325.1  &55.8   &\nodata        &14.3   &14.6\\
\hline

\enddata  
\tablenotetext{a} {The H$\beta$ flux is not available for this target. We give the H$\alpha$ flux instead. All line ratios are calculated with
respect to F$_{\rm H\alpha}=100$ for this PN.}
\tablenotetext{b}{This measurement include the H$\alpha$ and \nii blended fluxes.} 
\end{deluxetable}

\begin{deluxetable}{lrrclll}
\tabletypesize{\scriptsize}
\tablecaption {Positions, Dimensions and Morphologies of SMC Planetary Nebulae \label{Morph}}
\tablehead {
\colhead {} & \colhead {R. A.} & \colhead {Dec.} & \colhead {R$_{Phot}$} & 
\colhead {Dimensions} & \colhead {Morphological} & \colhead {} \\
\colhead {Nebula} & \colhead {(J2000)} & \colhead {(J2000)} & 
\colhead {(arcsec)} & \colhead {(arcsec)} & \colhead {Classification} & 
\colhead {Notes} 
}
\startdata  

J~4     &  0$^h$ 45$^m$ 27\fs30 &  $-73$\arcdeg 42\arcmin  15\farcs3 & 0.32    & 0.27        & E & Possible ansae. \\
J~18\tablenotemark{a}    &  0 51 43.39 &  $-73$ 00 54.1 & 0.17    & 0.14        & R? & Elongated in \nii. \\
J~23\tablenotemark{a}     &  0 55 30.52 &  $-72$ 50 21.3 & 0.38    & \nodata     & \nodata & Broad \ha emission; unresolved.\\
J~27\tablenotemark{a}    &  0 59 43.48 &  $-72$ 57 17.9 & \nodata & 2.5 x 1.7   & B? & very low surface brightness.\\
MA~1682\tablenotemark{a}  &  1 09 03.52 &  $-72$ 29 05.2 & 0.83    & 2.86 x 2.17 & B &  \\
MA~1762\tablenotemark{a}  &  1 12 40.28 &  $-72$ 53 46.4 & 0.71    & 1.45 x 1.26 & E(bc) &  \\
MG 8    &  0 56 19.59 &  $-72$ 06 58.5 & 0.48    & 1.39 x 1.28 & E & Inner structure. \\
MG~13\tablenotemark{a}   &  1 43 10.33 &  $-72$ 57 03.2 & 1.55    & 1.22 x 1.09 & E & Inner structure.\\
SMP~1   &  0 23 58.67 &  $-73$ 38 03.8 & 0.15    & \nodata     & \nodata & Unresolved. \\
SMP~6   &  0 41 27.52 &  $-73$ 47 06.2 & 0.19    & \nodata     & E &  \\
SMP~8   &  0 43 25.17 &  $-72$ 38 18.9 & 0.23    & 0.41 x 0.38 & R &  \\
SMP~9     &  0 45 20.66 &  $-73$ 24 10.5 & 0.55 & 1.20& R & inner structure. \\
SMP~11     &  0 48 36.61 &  $-72$ 58 00.1 & 0.99    & 0.78 x 0.66 & B &  possible inner ring.\\
SMP~12  &  0 49 21.00 &  $-73$ 52 58.0 & 0.37    & 0.78 x 0.51 & E &  \\
SMP~13    &  0 40 51.71 &  $-73$ 44 21.3 & 0.19    & 0.20        & R &  \\
SMP~14  &  0 50 34.99 &  $-73$ 42 57.9 & 0.42    & 0.83        & R & Ansae and/or inner structure.\\
SMP~17  &  0 51 56.41 &  $-71$ 24 44.6 & 0.25    & 0.50        & E & Faint detached 1\farcs5 halo; inner ring in \nii. \\
SMP~18    &  0 51 57.97 &  $-73$ 20 30.1 & 0.15    & 0.14        & \nodata &  unresolved. \\
SMP~19    &  0 53 11.14 &  $-72$ 45 07.5 & 0.30    & 0.59        & R &  outer structure. \\
SMP~20  &  0 56 05.39 &  $-70$ 19 24.7 & 0.15    & 0.20 x 0.23 & \nodata & Unresolved.\\
SMP~22  &  0 58 37.44 &  $-71$ 35 49.1 & 0.40    & 0.71 x 0.54 & B? & Possible ansae.\\
SMP~23    &  0 58 42.14 &  $-72$ 56 59.6 & 0.30    & 0.66 x 0.60 & E(bc)  &  \\
SMP~24  &  0 59 16.09 &  $-72$ 01 59.7 & 0.20    & 0.38        & E &  \\
SMP~25  &  5 30 33.22 &  $-70$ 44 38.4 & 0.19    & \nodata     & E &  \\
SMP~26  &  1 04 17.81 &  $-73$ 21 51.2 & 0.28    & 0.61 x 0.57 & P &  \\
SMP~27  &  1 21 10.67 &  $-73$ 14 35.4 & 0.23    & 0.45        & R & Attached outer halo.\\
SP~34   &  1 12 10.76 &  $-71$ 26 50.2 & 0.61    & 0.71 x 0.69 & R & Attached outer halo.\\
\enddata  
\tablenotetext{a}{No \oiii image available.}

\end{deluxetable}

\clearpage
\begin{deluxetable}{lrrr}
\tablecaption {PN Morphological Types: SMC versus LMC, and Galaxy}
\tablehead {
\colhead {Morphological Classification} & \colhead {SMC} & \colhead {LMC} & 
\colhead {Galaxy} \\
\colhead{}& \colhead{($\%$)} & \colhead{($\%$)} & 
\colhead {($\%$)}\\
}
\startdata
Round (R) &                             30 &    29&     23 \\
Elliptical (E) &                        37&     17&     49\\
Bipolar core (BC) &                     17 &    34&     9\\
Bipolar (B)\tablenotemark{a} &          13 &                    17&     17\\
Point-symmetric (P) &                   3 &                     3&      3\\
Total, Asymmetric\tablenotemark{b} &    30 &                    51&     26\\

\enddata  

\tablenotetext{a}{Includes Quadrupolar PNs}
\tablenotetext{b}{Includes B and BC, but not P}

\end{deluxetable}

\clearpage
\begin{deluxetable}{lrrrrrl}
\tabletypesize{\scriptsize}
\tablecaption {Photoionization Models, Input}
\tablehead {
\colhead {Model} & \colhead {log He/H} & 
\colhead {log C/H}& \colhead{log N/H}& \colhead{log O/H}& 
\colhead{log T$_{\rm eff}$}& \colhead{log L/L$_{\odot}$}\\
}
\startdata
------Galactic models &&&&&&\\
1&      -1.0&   -3.523& -4.0&   -3.222& 4.517&  3.887\\
2&      -1.0&   -3.523& -4.0&   -3.222& 4.828& 3.869\\
3&      -1.0&   -3.523& -4.0&   -3.222& 5.032& 3.823 \\
4&      -1.0&   -3.523& -4.0&   -3.222&  5.202& 3.655 \\
5&      -1.0&   -3.523& -4.0&   -3.222&  5.238& 3.348\\
------LMC models &&&&&&\\
6&      -1.0&   -3.380& -3.91&  -3.644& 4.569&  3.973\\ 
7&      -1.0&   -3.380& -3.91&  -3.644& 4.797&   3.964 \\
8&      -1.0&   -3.380& -3.91&  -3.644& 5.024&   3.932\\
9&      -1.0&   -3.380& -3.91&  -3.644& 5.233&   3.805  \\
10&     -1.0&   -3.380& -3.91&  -3.644&  5.298&   3.481\\
------SMC models &&&&&&\\
11&     -1.03&  -3.370& -4.57&  -4.06&  4.523&   3.938\\ 
12&     -1.03&  -3.370& -4.57&  -4.06&  4.838&   3.936\\
13&     -1.03&  -3.370& -4.57&  -4.06&  5.046&   3.905\\
14&     -1.03&  -3.370& -4.57&  -4.06&  5.234&   3.775\\
15&     -1.03&  -3.370& -4.57&  -4.06&  5.292&   3.474\\

\enddata

\end{deluxetable}

\begin{deluxetable}{lrr}
\tabletypesize{\scriptsize}
\tablewidth{200pt}
\tablecaption {Electron Density and Ionized Mass}
\tablehead {
\colhead {name} & \colhead {log N$_{\rm e}$} & \colhead {M${\rm ion}$} 
 \\
\colhead{}& \colhead{cm$^{-3}$} & \colhead{\sm} \\
}
\startdata
SMC&&\\

      SMP~13 &        3.6&      0.40  \\    
      SMP~14 &      3.6&        0.10\\
      SMP~18&       3.6&        0.28  \\
      SMP~19&      3.4& 0.21\\
      SMP~22 &      3.4&        0.28    \\      
      SMP~24 &      3.1&        0.86    \\
      SMP~25 &      3.4&        0.10\\
      SMP~26 &      3.1&        0.15 \\
      
      average&  3.4&    0.30\\
 
 LMC&&\\
       SMP~9 &      3.2     & 0.12\\
      SMP~13  &      2.9     & 0.63\\
    SMP~16\tablenotemark{a}   &    2.9      & 0.33\\
    SMP~19    &   3.4      & 0.21\\
      SMP~28  &     3.2     &  0.10\\
    SMP~30\tablenotemark{a}   &    2.9      &  0.16\\
    SMP~46    &   3.4      & 0.05\\
     SMP~53   &     3.5      & 0.26\\
      SMP~71  &     3.6      & 0.17\\
      SMP-80  &     3.2      & 0.14\\
     SMP~95\tablenotemark{a}  &     2.9      & 0.16\\
    SMP~100   &    3.4    & 0.17\\
average\tablenotemark{b}&       3.3 &   0.21\\
 
\enddata  

\tablenotetext{a} {Possible overalp of the \sii 6716 and 6731 lines,
uncertain electron density and ionized mass.}
\tablenotetext{b} {do not include PNs with uncertain ratios (see $^a$).}

\end{deluxetable}

\end{document}